\documentclass[
    reprint,
    superscriptaddress,
    nofootinbib,
    amsmath,
    amssymb,
    aps,
    prl,
    floatfix,
    longbibliography
]{revtex4-2}
    
    \usepackage{graphicx}
    \usepackage[caption=false]{subfig}
    \usepackage[usenames,dvipsnames]{xcolor} 
    \usepackage{pgfplots}
    \pgfplotsset{compat=1.18}
    \usepackage[utf8]{inputenc}
    \usepackage{color}
    \usepackage{hyperref}
    \usepackage[normalem]{ulem} 
    \usepackage{physics}
    \usepackage{enumitem}
    \usepackage{comment}
    \usepackage{bm}
    \usepackage{aas_macros}
    \usepackage{appendix}
    \usepackage{orcidlink}
\hypersetup{
  breaklinks = true,
  colorlinks   = true, 
  urlcolor     = blue, 
  linkcolor    = blue, 
  citecolor   = blue 
}

\usepackage{amsmath,amssymb,bm}
\usepackage{graphicx}
\usepackage{booktabs}
\usepackage{microtype}
\usepackage{xcolor}
\usepackage{hyperref}
\hypersetup{colorlinks=true,citecolor=blue,linkcolor=blue,urlcolor=blue,pdfauthor={Susana Landau and Jailson S. Alcaniz and Agripino Sousa-Neto},pdftitle={Testing the Robustness of Dynamical Dark Energy with Cosmic Age}}

\newif\ifshowcomments
\showcommentstrue

\definecolor{SousaColor}{RGB}{0,102,204}
\definecolor{JailsonColor}{RGB}{204,51,51}
\definecolor{SusanaColor}{RGB}{130,60,180}

\begin{document}
\title{Are Dynamical Dark-Energy Cosmologies Old Enough?}

\author{Agripino Sousa-Neto}\email{agripinoneto@on.br}
\affiliation{Observat\'orio Nacional, 20921-400 Rio de Janeiro, RJ, Brazil}
\author{Susana J. Landau}\email{slandau@df.uba.ar}
\affiliation{Universidad de Buenos Aires, Facultad de Ciencias Exactas y Naturales, Departamento de Física, Ciudad Universitaria 1428,
Buenos Aires, 1460, Argentina}
\affiliation{CONICET - Universidad de Buenos Aires, Instituto de Física de Buenos Aires (IFIBA), Ciudad Universitaria 1428, Buenos Aires,
1460, Argentina}
\affiliation{Observat\'orio Nacional, 20921-400 Rio de Janeiro, RJ, Brazil}
\author{Jailson Alcaniz}
\email{alcaniz@on.br}
\affiliation{Observat\'orio Nacional, 20921-400 Rio de Janeiro, RJ, Brazil}

\date{\today}

\begin{abstract}
Recent analysis combining baryon acoustic oscillation measurements with cosmic microwave background and Type Ia supernova data show an apparent preference for a time-dependent dark-energy equation of state within $w_0w_a$ parameterizations. Here we test the robustness of this result against an independent cosmological consistency requirement, namely that the expansion history must provide sufficient time to accommodate the oldest observed stellar populations. Motivated by recent stellar and globular-cluster age determinations, we impose progressively stronger lower bounds, $t_0 \geq 13.6$, $13.75$, $13.8$, and $13.85$ Gyr, on the expansion histories allowed by DESI DR2+CMB+Pantheon+ data. We find a systematic displacement of the posterior in the $(w_0,w_a)$ plane toward the $\Lambda$CDM prediction, $(w_0,w_a)=(-1,0)$, as the minimum allowed age is increased. For $t_0 \geq 13.85$ Gyr, the Mahalanobis separation from $\Lambda$CDM decreases from $2.56\sigma$ to $1.72\sigma$, bringing the $\Lambda$CDM prediction inside the 95\% credible region. Our results show that the current evidence for dynamical dark energy is sensitive to independent cosmic-age information and that stellar chronometry provides a direct robustness test of the DESI-favored expansion histories. Requiring sufficiently old cosmologies can appreciably reduce the statistical preference for departures from the standard $\Lambda$CDM model.

\end{abstract}

\maketitle

\label{sec:introduction}

{\emph{Introduction —}} The age of the Universe has long provided a fundamental consistency test of cosmological models: a viable expansion history must allow sufficient time for the formation and evolution of the oldest observed objects~\cite{Sandage1961}. This consistency requirement was put to the test soon after the discovery of cosmic expansion, when the large values of the Hubble constant, $H_0$, inferred in the 1930s implied an expansion age shorter than contemporary estimates of the age of the Earth, giving rise to the first cosmological time-scale problem~\cite{North1965}. Several decades later, the problem resurfaced in the ``age crisis"  of the 1990s, when estimates for the oldest Galactic globular clusters, reaching $\gtrsim 14$ Gyr, were difficult to reconcile with the expansion age of a spatially flat, matter-dominated Universe~\cite{BolteHogan1995}. Cosmic-age considerations consequently provided a fundamental argument for a nonzero cosmological constant and, more generally, for cosmic acceleration~\cite{KraussTurner1995,OstrikerSteinhardt1995,AlcanizLima1999}.

The current situation is qualitatively different, but the underlying consistency argument remains unchanged. Recent advances in stellar chronometry have renewed the role of cosmic age as an independent cosmological probe. In particular, Ref.~\cite{Valcin2026} analyzes high-quality \textit{Hubble Space Telescope} color--magnitude diagrams of 69 Galactic globular clusters, explicitly allowing for multiple stellar populations with independent ages, metallicities, helium abundances, and population fractions within a hierarchical inference framework. For the metal-poor subsample, they infer an age of $13.61\pm0.25,({\rm stat})\pm0.23,({\rm sys})$ Gyr for the dominant old population, which, after accounting for the time delay between the Big Bang and globular cluster formation, yields $t_0=13.81\pm0.25,({\rm stat})\pm0.23,({\rm sys})$ Gyr (see also \cite{Valcin_2025}). The authors of Ref.~\cite{Banik2026} instead use a large sample of Galactic field stars selected from LAMOST spectroscopy and \textit{Gaia} astrometry. Combining isochrone-based stellar ages with metallicity and $\alpha$-enhancement criteria and reconstructing the underlying age distribution through an MCMC analysis, they infer $A_\star=13.73^{+0.18}_{-0.15}$ Gyr for the oldest stellar population, with estimates ranging from $13.31^{+0.21}_{-0.18}$ to $14.02^{+0.18}_{-0.15}$ Gyr under different selection criteria. Complementary constraints from old main-sequence turn-off and subgiant stars yield similarly old cosmic ages \cite{Tomasetti:2025jpg}, reinforcing the consistency of these independent stellar chronometers.

At the same time, baryon acoustic oscillation (BAO) measurements from the Dark Energy Spectroscopic Instrument (DESI), when combined with cosmic microwave
background (CMB) and type Ia supernovae (SN) observations, favor regions of the
 $w_0w_a$ parameter space that depart from the $\Lambda$CDM prediction, with a significance of $2.8$--$4.2\sigma$, depending on the adopted supernova compilation~\cite{DESI:2025zgx}. If confirmed, such a result would have profound implications for cosmology, challenging the simplest interpretation of cosmic acceleration and sharpening the search for its underlying physical origin. However, constructing a physically motivated model that reproduces the favored evolution remains challenging. In particular, the preferred dark-energy
histories typically involve a crossing of the phantom divide ($w=-1$), which cannot be realized by a conventional, minimally coupled single scalar field~\cite{2002PhLB..545...23C}. A viable explanation may therefore require additional ingredients, such as multiple fields or interactions in the dark sector~\cite{2010PhR...493....1C,deSouza:2026qyf,Babu:2026tfr}.

Thus, given the historical role of cosmic-age inconsistencies in challenging otherwise viable cosmological models, this possible departure from $\Lambda$CDM suggested by DESI naturally raises a fundamental consistency question: do the favored expansion histories allow sufficient cosmic time to accommodate the oldest independently dated stellar populations?

\begin{figure*}[t]
    \centering
    \includegraphics[width=\textwidth]{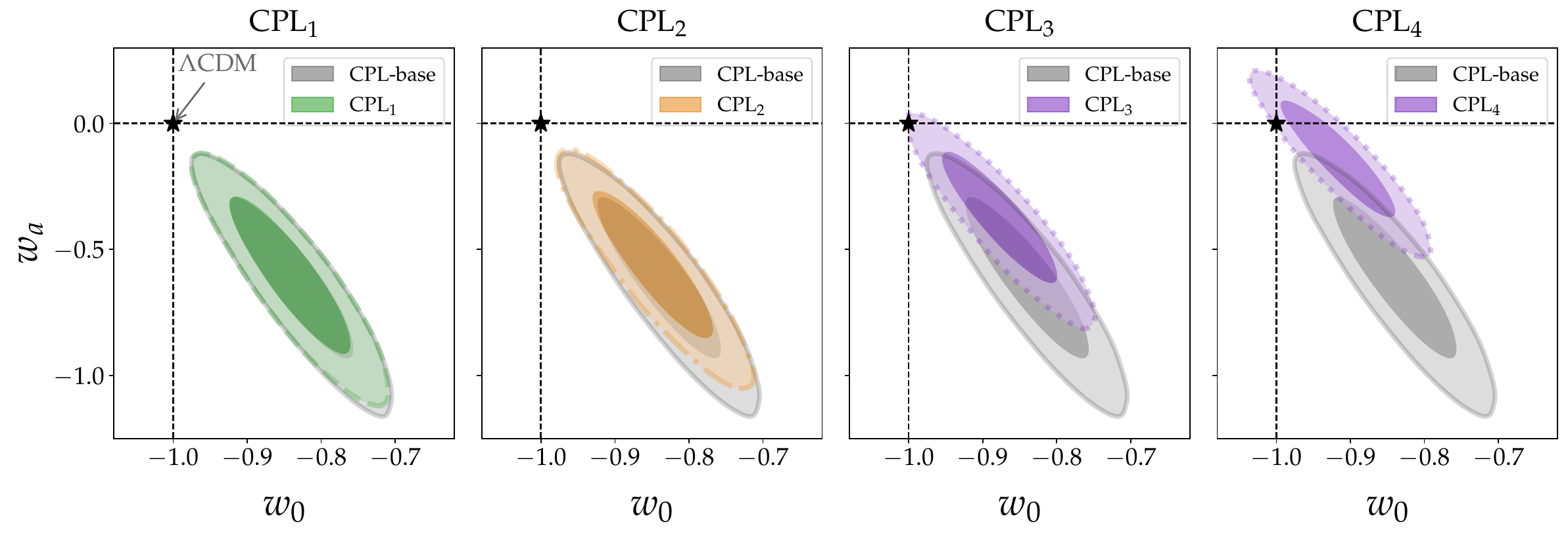}
   \caption{Marginalized constraints in the $(w_0,w_a)$ plane for the 
CPL-base and $\rm{CPL_{1 - 4}}$ scenarios. The star marks the $\Lambda$CDM prediction,
$(w_0,w_a)=(-1,0)$.}
    \label{fig:w0wa_age}
\end{figure*}

In this work, we answer this question by testing the robustness of the apparent preference for dynamical dark energy (DDE) against the independent cosmic-age requirement discussed above. Using DESI DR2 BAO combined with CMB and SN data, we impose progressively stronger lower bounds on the cosmic age, motivated by current age estimates from globular clusters and the oldest stars~\cite{Valcin2026}. We find that requiring older cosmic histories systematically shifts the inferred dark-energy parameters toward the standard $\Lambda$CDM prediction. For example, imposing a lower age bound of $t_0 \geq 13.85$ Gyr reduces the statistical separation from $\Lambda$CDM to $1.72\sigma$, bringing the standard model within the 95\% credible region. Our results therefore show that the inferred preference for DDE is sensitive to independent constraints on the age of the Universe, providing a direct test of the consistency of the DESI-favored expansion histories with stellar chronometry.

{\emph{Cosmic age, data and methodology —}} For a spatially flat Universe, the age of the Universe is determined by the expansion history according to
\begin{equation}
\label{eq}
t_0=\int_0^{t_0} {dt} 
=\frac{1}{H_0}\int_0^1 \frac{da}{aE(a)},
\end{equation}

where $E(a)\equiv H(a)/H_0$ and the scale factor is written in terms of redshift $z$ as $a = (1+z)^{-1}$. The CPL parametrization for the dark-energy equation of state is given by 
$w(a)=w_0+w_a(1-a)$ or, equivalently, $w(z)=w_0+w_a z/(1+z)$. Thus, neglecting radiation at late times, the corresponding expansion rate is
\begin{equation}
\label{eqE}
E^2(a) = \Omega_{m}a^{-3} + (1-\Omega_{m}) a^{-3\lambda} \exp\left[3w_a(a-1)\right]\;,
\end{equation}

where $\lambda = 1+w_0+w_a$ and $\Omega_m$ corresponds to the present-day matter density parameter. The radiation contribution, $\Omega_ra^{-4}$,  can straightforwardly be included in Eq.~(\ref{eqE}) and is retained in our numerical analysis.

The cosmic age therefore depends directly on the dark-energy parameters through their effect on the integrated expansion history. At fixed values of $H_0$ and $\Omega_m$, increasing $w_0$ above $-1$ generally increases the dark energy density in the past and hence the expansion rate, reducing $t_0$, whereas more negative values of $w_0$ tend to produce older cosmic histories. The parameter $w_a$ controls the redshift evolution of the equation of state: positive $w_a$ typically enhances the dark energy contribution at earlier times and decreases the age, while negative $w_a$ tends to suppress it and increase $t_0$. Because the region favored by current data combines $w_0>-1$ with $w_a<0$ ($
w_0=-0.838\pm0.055; 
w_a=-0.62^{+0.22}_{-0.19}$ at 68\% c.l.)~\cite{DESI:2025zgx}, its effect on $t_0$ is determined by the joint evolution of the two parameters and by their correlations with the remaining cosmological parameters. We therefore assess the impact of cosmic-age information directly within the full parameter inference described below.

We use the combination of CMB, BAO, and SN Ia data for which recent DESI analyses have reported a preference for DDE over $\Lambda$CDM. We use the Planck 2018 low-$\ell$ TT and EE likelihoods \cite{Planck:2018vyg}, together with the Planck NPIPE high-$\ell$ CamSpec TT, TE, and EE likelihood and the ACT DR6 CMB-lensing likelihood \cite{AtacamaCosmologyTelescope:2025blo} in its ACT+Planck baseline configuration. We combine these measurements with DESI DR2 BAO data \cite{DESI:2025zgx} and the Pantheon+ SN Ia compilation \cite{Scolnic_2022}, restricted to $z>0.01$.

We refer to the CPL analysis with this data combination and no cosmic-age requirement as {CPL-base}, which reproduces the DESI DR2 results and serves as the reference case throughout this work. To assess the sensitivity of the DDE  constraints to the cosmic age, we define four additional scenarios, $\mathrm{CPL}_1$--$\mathrm{CPL}_4$, by supplementing the CPL-base with progressively stronger lower bounds on the age of the Universe, $t_{\min}=13.60,\ 13.75,\ 13.80,$ and $13.85$ Gyr, respectively. In all cases, the cosmological model and data combination are otherwise identical to CPL-base. The age requirement is implemented as the hard prior
\begin{equation}
\ln {\cal L}_{\rm age} =
\begin{cases}
0,       & t_0(\boldsymbol{\theta}) \geq t_{\rm min},\\
-\infty, & t_0(\boldsymbol{\theta}) < t_{\rm min},
\end{cases}
\label{eq:age_prior}
\end{equation}
with the total likelihood given by
\[
\ln {\cal L}_{\rm tot}
=
\ln {\cal L}_{\rm Planck}
+\ln {\cal L}_{\rm BAO}
+\ln {\cal L}_{\rm SN}
+\ln {\cal L}_{\rm age}.
\]
These hard cuts are not intended to represent the full likelihood of any specific stellar-age determination, but rather to assess the robustness of the apparent evidence for DDE obtained within the CPL-base model as progressively older cosmic histories are required.

\emph{Results —} Fig.~\ref{fig:w0wa_age} shows the central result of our analysis. CPL-base reproduces the DESI DR2 analysis and favors the familiar elongated region in the $(w_0,w_a)$ plane away from the $\Lambda$CDM model. However, from $\mathrm{CPL}_1$ to $\mathrm{CPL}_4$, a clear trend emerges: as the minimum allowed age is increased, the posterior shifts progressively toward smaller values of $w_0$ and less negative values of $w_a$, and hence toward the $\Lambda$CDM point, $(w_0,w_a)=(-1,0)$, while the portion of parameter space associated with the largest departure from $\Lambda$CDM is progressively suppressed. 

This trend is not specific to any single age threshold; rather, it shows a continuous relation between the minimum cosmic age allowed in the analysis and the inferred departure from $\Lambda$CDM. Expansion histories lying furthest along the dynamical-dark-energy direction favored by the baseline likelihood tend to predict younger universes, whereas imposing stronger age requirements preferentially retains models closer to $\Lambda$CDM. Although not shown here, a similar result is obtained using the DES-Dovekie compilation, an updated reanalysis of the five-year Dark Energy Survey supernova sample~\cite{DES:2025sig}.

\begin{figure}[t]
    \centering
    \includegraphics[width=0.80\linewidth]{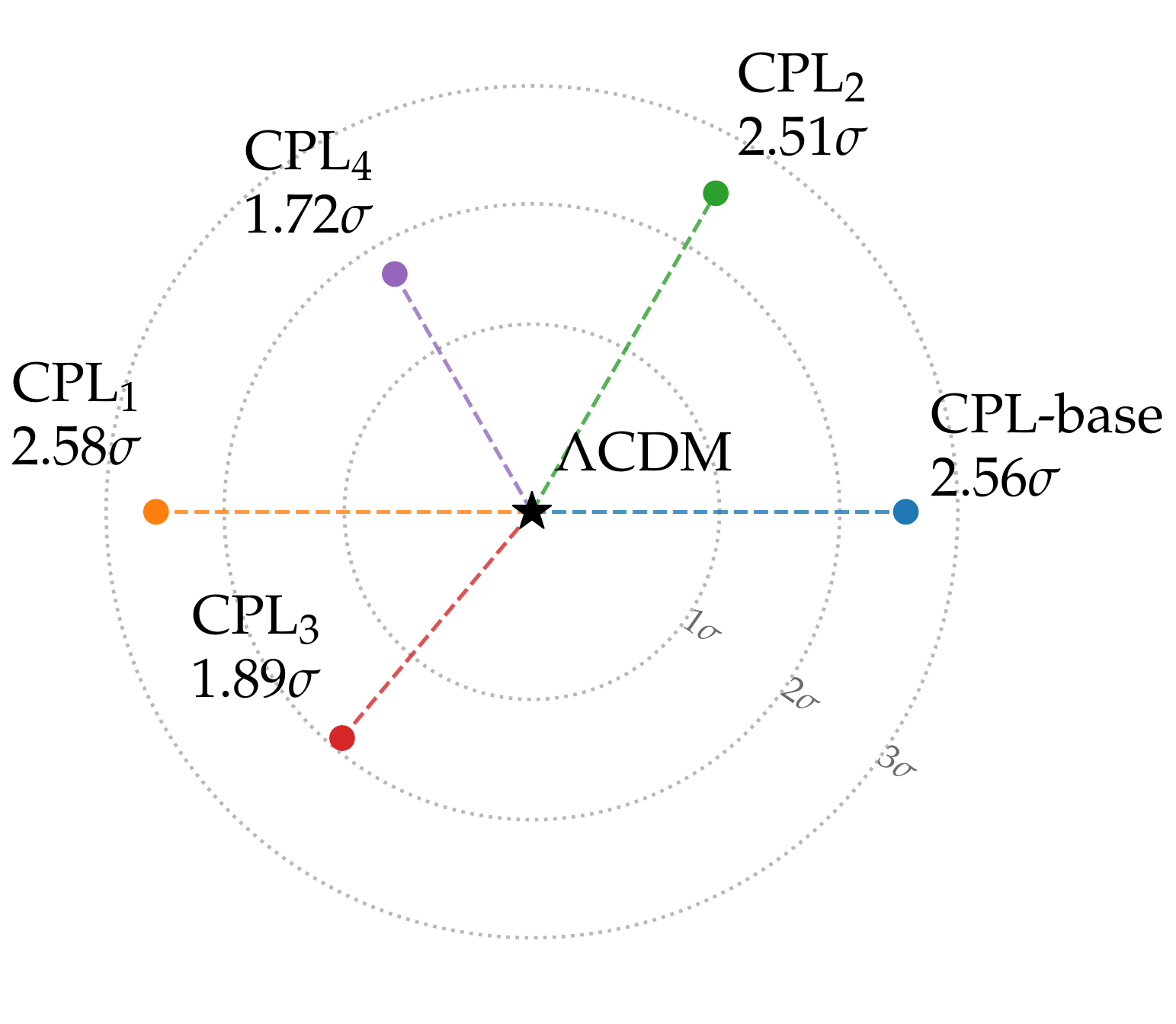}
  \caption{Covariance-weighted separation from the $\Lambda$CDM
prediction for CPL-base and $\rm{CPL_{1 - 4}}$ scenarios.
The radial coordinate represents $\sqrt{D_{\rm Maha}}$, while
the dotted circles indicate the equivalent Gaussian significance.
Angular positions are arbitrary and chosen only for visual clarity.}
    \label{fig:D_circle}
\end{figure}

We quantify the departure from $\Lambda$CDM using the generalized Mahalanobis distance of the CPL posterior, 
\begin{equation}
D_{\rm Maha}
=
(\hat{\boldsymbol{\theta}}-\boldsymbol{\theta}_{\Lambda{\rm CDM}})^T
C^{-1}
(\hat{\boldsymbol{\theta}}-\boldsymbol{\theta}_{\Lambda{\rm CDM}}),
\label{eq:mahalanobis}
\end{equation}
where $\hat{\boldsymbol{\theta}}=(w_0,w_a)$ is the posterior mean,
$\boldsymbol{\theta}_{\Lambda{\rm CDM}}=(-1,0)$, and $C$ is the
corresponding covariance matrix. Under the Gaussian approximation,
$D_{\rm Maha}$ follows a $\chi^2$ distribution with two degrees of
freedom, from which we obtain the equivalent two-sided Gaussian
significance, $N_\sigma$. For the baseline case, we also compute
$\Delta\chi^2$, obtaining $\Delta\chi^2=10.73$ ($2.83\sigma$),
consistent with the preference reported by DESI~\cite{DESI:2025zgx}. It is important to mention that the two significances need not coincide: while $\Delta\chi^2$ compares the
best fits from separate CPL and $\Lambda$CDM analyses, 
$D_{\rm Maha}$ quantifies the separation of the fixed $\Lambda$CDM
point $(-1,0)$ from the CPL posterior using its covariance.

\begin{figure}[t]
\includegraphics[width=0.99\linewidth]{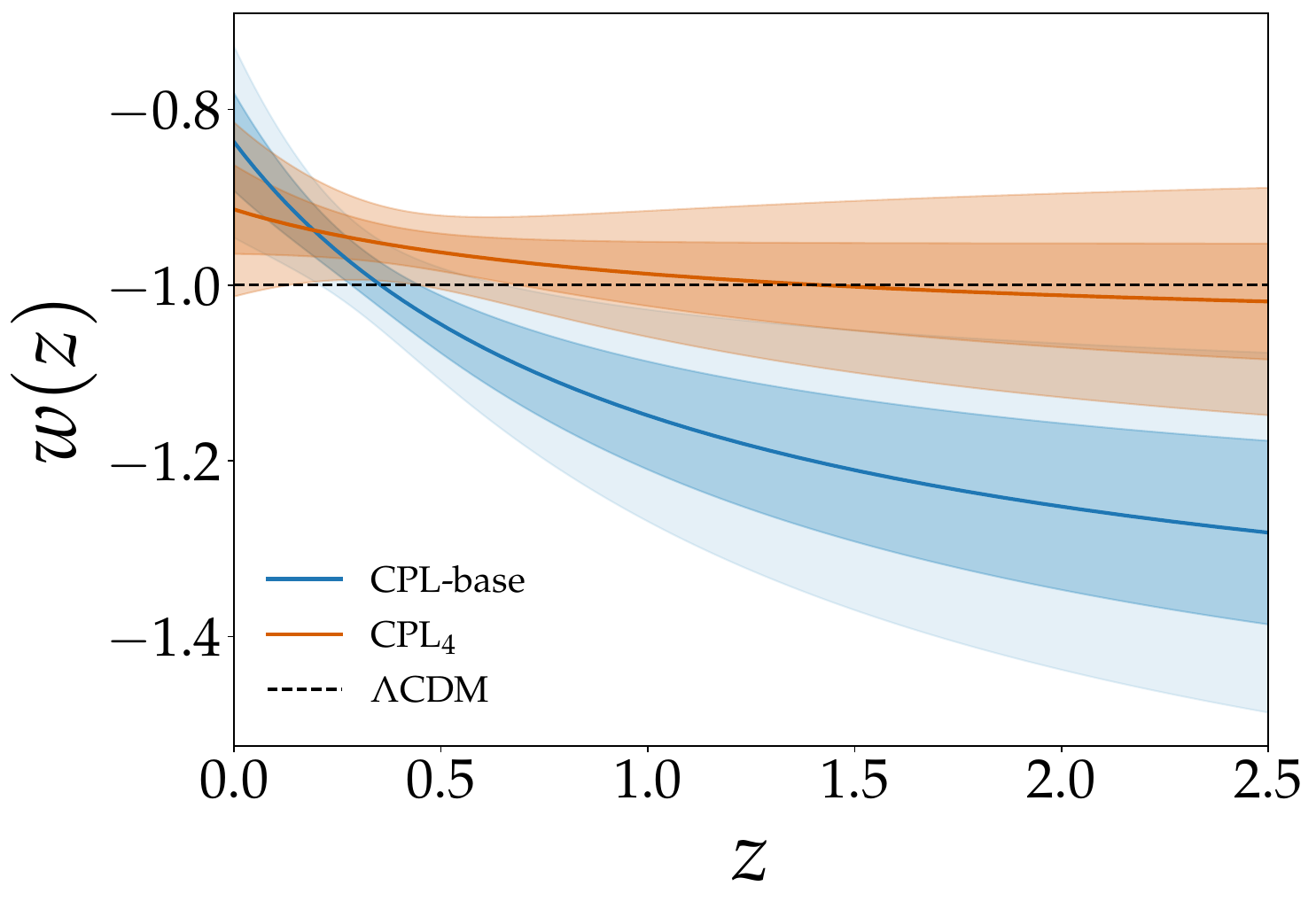}
\caption{Posterior mean and 68\% and 95\% credible regions
for $w(z)$ for CPL-base and CPL$_4$. The cosmic-age prior
in CPL$_4$ shifts the reconstruction toward the $\Lambda$CDM
prediction.}
\label{fig:wz}
\end{figure}

The corresponding separation from $\Lambda$CDM is reported in 
Fig.~2. It remains essentially unchanged from CPL-base ($2.56\sigma$) to $\mathrm{CPL}_1$ ($2.58\sigma$) and $\mathrm{CPL}_2$ ($2.51\sigma$), but decreases to $1.89\sigma$ for $\mathrm{CPL}_3$ and $1.72\sigma$ for $\mathrm{CPL}_4$, bringing the $\Lambda$CDM prediction  within the $95\%$ credible region in the latter two cases. The origin of this shift can also be seen directly from the marginalized age distributions. The baseline CPL posterior extends toward younger cosmic ages, while successively stronger age requirements progressively remove this region of parameter space. Because $t_0$ is derived from an integral over $H^{-1}$, these cuts propagate into the joint $w_0$--$w_a$ posterior rather than constraining either parameter independently.

For completeness, we also reconstruct the evolution of the dark energy equation-of-state parameter, shown in Fig.~3, which provides an alternative view of the effect discussed above. Without the age requirement, the preferred CPL region typically satisfies $w_0>-1$ and $w_a<0$, implying a phantom-like equation of state at intermediate redshift, followed by a crossing of $w=-1$ and an evolution toward $w>-1$ at late times. Such a crossing can be difficult to realize in simple physically motivated dark energy models, providing additional motivation to examine whether this behavior remains robust under independent consistency requirements. Restricting the analysis to sufficiently old cosmologies substantially weakens this evolution and shifts $w(z)$ toward the $\Lambda$ prediction throughout the redshift range constrained by the data. The cosmic-age requirement therefore not only reduces the available parameter volume, but also preferentially suppresses the expansion histories associated with the strongest  preference for DDE.


\emph{Discussion and conclusions —} Cosmic-age considerations have played a recurring role in testing the internal consistency of cosmological models, particularly when new observational results pointed beyond the prevailing paradigm. In this work, we applied this basic requirement to the DDE  cosmologies currently favored by combinations of DESI DR2 BAO, CMB, and SN Ia observations.

Our main result is a systematic relation between the minimum cosmic age allowed by the analysis and the inferred departure from $\Lambda$CDM. As progressively stronger lower bounds on $t_0$ are imposed, the allowed region in the $(w_0,w_a)$ plane shifts toward the $\Lambda$CDM prediction. For $t_0>13.8$ Gyr, motivated by recent stellar chronometry, the covariance-weighted separation from $\Lambda$CDM decreases from $2.56\sigma$ to $1.89\sigma$, bringing $(w_0,w_a)=(-1,0)$ within the $95\%$ credible region; for $t_0>13.85$ Gyr, the separation is further reduced to $1.72\sigma$. The apparent preference for dynamical dark energy is therefore closely associated with expansion histories that predict comparatively younger universes.

The origin of this behavior is straightforward. Since the cosmic age depends on the integral of $H^{-1}$ over the entire expansion history, it constrains a combination of $w_0$ and $w_a$ that is complementary to the geometrical information provided by BAO, CMB, and SN data. This behavior shows that the cosmic-age information selectively disfavors those CPL histories that depart most strongly from $\Lambda$CDM, rather than simply shrinking the allowed parameter space. The corresponding reconstruction of $w(z)$ shows the same trend, with increasingly stringent age requirements weakening the characteristic evolution favored by the baseline CPL analysis and shifting the equation of state toward $w=-1$ over the redshift range constrained by the data.

It is important to emphasize that the hard age cuts adopted here should not be interpreted as full likelihoods for individual stellar-age determinations. Current estimates of the ages of globular clusters and the oldest stars carry both statistical and systematic uncertainties, and our sequence of lower bounds is intended instead as a controlled sensitivity test of the CPL-base constraints.

Finally, the historical role of cosmic-age arguments suggests that proposed departures from $\Lambda$CDM should be tested not only by their improvement in the fit to cosmological data, but also by the physical consistency of the resulting expansion histories. Our results show that this consistency requirement has a significant impact, in that sufficiently old cosmic histories substantially reduce the inferred departure from $\Lambda$CDM. In this sense, stellar chronometry provides a simple and independent check on the robustness of current claims for evolving dark energy.

{\emph{Acknowledgments —}} ASN acknowledges support from the Conselho Nacional de Desenvolvimento Cient\'{\i}fico e Tecnol\'ogico (CNPq) through the PCI/Observat\'orio Nacional grant No.~402113/2026-8. JSA is supported by CNPq grants No.~307683/2022-2 and 448158/2025-6, and from the Funda\c{c}\~ao de Amparo \`a Pesquisa do Estado do Rio de Janeiro (FAPERJ) under grant No.~299312 (2023).
\bibliography{age_DDE}
\end{document}